\documentclass[10pt,letterpaper]{article}

\usepackage{amsmath,amssymb}

\usepackage{rotating,epsfig,graphicx}
\usepackage{times}
\usepackage{cite}

\date{}
\usepackage{lastpage,fancyhdr,graphicx}
\usepackage{epstopdf}

\begin{document}

\begin{flushleft}
{\Large
\textbf\newline{Influenza Evolution and H3N2 Vaccine Effectiveness,
with Application to the 2014/2015 Season}
}
\newline

Xi Li\textsuperscript{1},
Michael W.\ Deem\textsuperscript{1,2,3,*}
\\
\bigskip
\bf{1} Department of Bioengineering, Rice University, Houston, TX 77005\\
\bf{2} Department of Physics \& Astronomy, Rice University, Houston, TX 77005\\
\bf{3} Center for Theoretical Biological Physics, 
       Rice University, Houston, TX 77005\\

* mwdeem@rice.edu
\end{flushleft}

\section*{abstract}

Influenza A is a serious disease that causes significant morbidity and mortality, and vaccines against the seasonal influenza disease are of variable effectiveness. In this paper, we discuss use of the 
 $p_{\rm epitope}$ 
method to predict the dominant influenza strain and the expected vaccine effectiveness in the coming flu season. We illustrate how the effectiveness of the 2014/2015 A/Texas/50/2012 [clade 3C.1] vaccine against the A/California/02/2014 [clade 3C.3a] strain that emerged in the population can be estimated via pepitope. In addition, we show by a multidimensional scaling analysis of data collected through 2014, the emergence of a new A/New Mexico/11/2014-like cluster [clade 3C.2a] that is immunologically distinct from the A/California/02/2014-like strains.

\section*{Author Summary}
We show that the $p_{\rm epitope}$ measure of antigenic distance
is correlated with influenza A H3N2 vaccine effectiveness in humans
with $R^2 = 0.75$ in the years 1971--2015.
As an example,
we use this measure to predict from sequence data prior to 2014 the effectiveness
of the 2014/2015 influenza vaccine against the
A/California/02/2014  strain that emerged in 2014/2015.
Additionally, we use this measure along with a reconstruction of the
probability density of the virus in sequence space from sequence
data prior to 2015 to predict that
a newly emerging A/New Mexico/11/2014 cluster will likely
be the dominant circulating strain in 2015/2016.

\section*{Introduction}

Influenza is a highly contagious virus, usually spread by 
droplet or fomite transmission.  The high mutation and reassortment
rates of this virus lead to significant viral diversity in the population
\cite{ferguson2003,Holmes}.
In most years, one type of influenza predominates among
infected people, typically A/H1N1, A/H3N2, or B.
In the 2014/2015 season, A/H3N2 was the most common \cite{Y14}.
While there are many strains of influenza A/H3N2, 
typically there is a dominant cluster of strains that infect
most people during one winter season.
Global travel by infected individuals leads this cluster of sequences
to dominate in most affected countries in a single influenza season.
New clusters arise every 3--5 years by the combined effects of
mutation and selection \cite{smith,clustering}. 
%Novel strains often originate in
%East or Southeast Asia in birds, passing through swine, and spreading
%to humans \cite{Russell2008}.
There is significant selection pressure upon the virus to evolve
due to prior vaccination or exposure \cite{Illingworth,Lassig}.
%Additionally, antiviral usage leads to evolution of drug resistance
%\cite{Foll}.

Due to evolution of the influenza virus, the strains selected
by the World Health Organization (WHO) for inclusion 
in the seasonal vaccine are reviewed annually and often updated.
The selection is based on which strains are circulating,
the geographic spread of circulating strains, and
the expected effectiveness of the current vaccine strains
against newly identified strains  \cite{CDC_flu}.
There are to date 143 national influenza centers
located in 113 countries that provide and study influenza
surveillance data.  
Five WHO Collaborating Centers for Reference and Research on Influenza 
(Centers for Disease Control and Prevention in Atlanta, Georgia, USA;
National Institute for Medical Research in London, United Kingdom; 
Victorian Infectious Diseases Reference Laboratory in Melbourne, Australia;
National Institute of Infectious Diseases in Tokyo, Japan; and
Chinese Center for Disease Control and Prevention in Beijing, China)
are sent samples for additional analysis.
These surveillance data are used to make forecasts about which
strains are mostly likely to dominate in the human population.
These forecasts are used by the WHO to make specific recommendations
about
the strains to include in the annual vaccine, 
 in 2016 one each of a A/H1N1, A/H3N2, and influenza B Yamagata lineage or Victoria lineage subtype strain. Additionally, for each recommended strain there is often a list of 5--6 ``like'' strains that may be substituted by manufacturers for the recommended strain and which may grow more readily in the vaccine manufacturing process that uses hen's eggs.

We here focus on predicting the
expected effectiveness of the current vaccine strains
against newly identified strains and on predicting or detecting
the emergence of new influenza strains.
Predicting effectiveness or emergence without recourse to
animal models or human data is challenging.
The influenza vaccine protects against strains similar to the vaccine, but not
against strains sufficiently dissimilar.  For example, the A/Texas/50/2012(H3N2)
2014/2015 Northern hemisphere
vaccine has been observed to not protect against the
A/California/02/2014(H3N2)  virus.
Furthermore, there is no vaccine that provides long-lasting, 
universal protection, although this is an active research 
topic \cite{universal}.
%We here show a mathematical way
%to predict from sequence data alone expected
%vaccine effectiveness \cite{calculator,gupta,flu2,h1n1,entropy}.
% We illustrate the method
%with an application to the 2014/2015 season.
%We show that this approach correlates with H3N2 vaccine
%effectiveness in humans with $R^2 = 0.75$.  

\begin{figure}[tb!]
\begin{center}
\includegraphics[width=0.45\columnwidth,clip=]{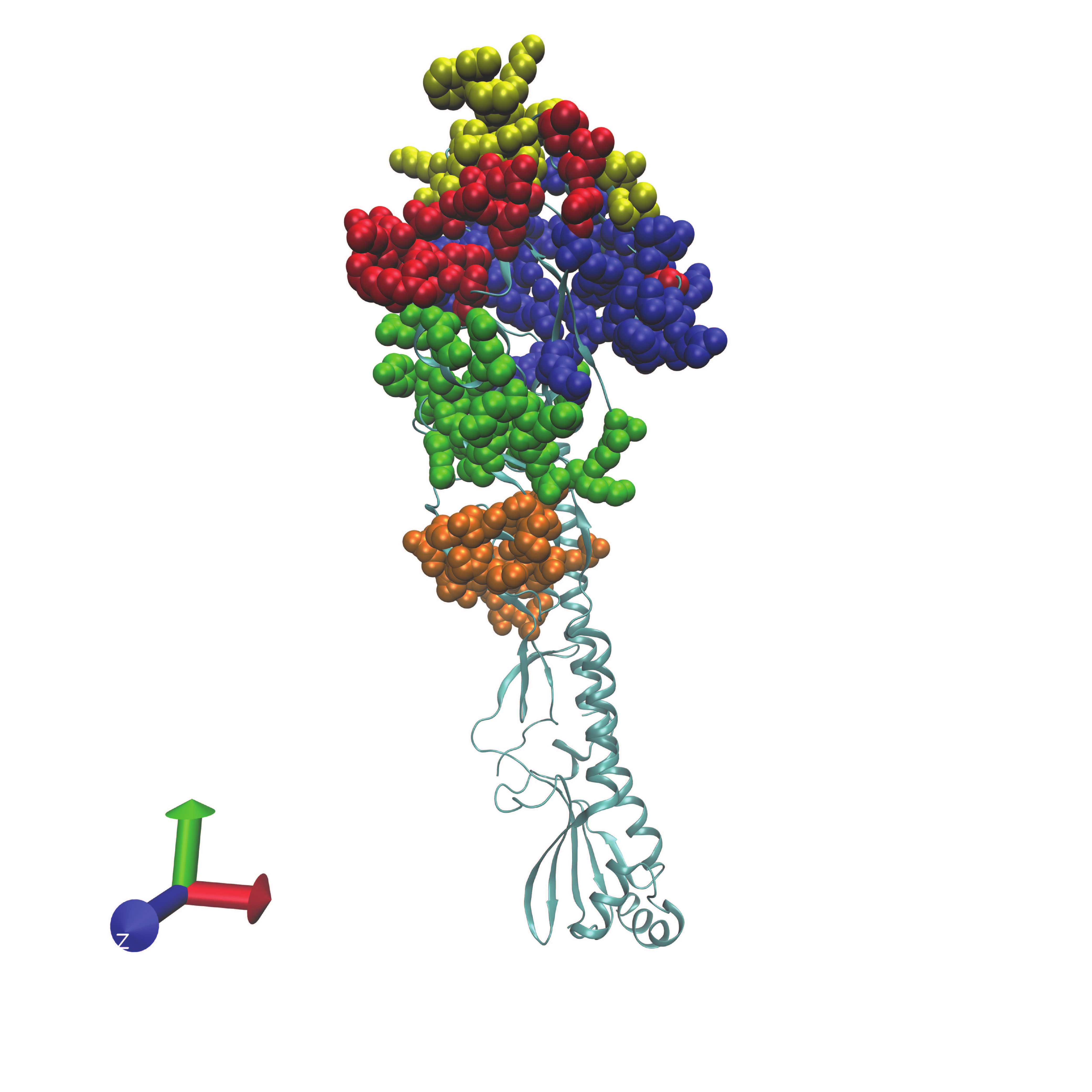}
\caption{
Shown is the structure of
hemagglutinin in H3N2 (accession number 4O5N).
The five epitope regions \cite{gupta} are color coded:
epitope A is red (19 amino acids), B is yellow (21 aa), C is orange (27 aa), D is blue
(41 aa), and E is green (22 aa).
Note epitope B was dominant in 2013/2014 and 2014/2015.
\label{fig0}
}
\end{center}
\end{figure}

Vaccine effectiveness is expected to be a function of
``antigenic distance.'' 
While antigenic distance is often estimated from
ferret animal model hemagglutination inhibition (HI)
studies, the concept is more general. 
In particular, in the present study we are interested in 
the antigenic distance that the human immune system detects.
A measurement of antigenic distance that is
predictive of vaccine effectiveness for H3N2 and H1N1 influenza A
in humans is $p_{\rm epitope}$ \cite{calculator,gupta,flu2,h1n1,entropy}.
%A theory of the immune system
%has proven useful in quantifying this selection pressure on the
%virus \left\cite{flu2,h1n1,entropy}.
The quantity $p_{\rm epitope}$ is the fraction of 
amino acids in the dominant epitope region of hemagglutinin that
differ between the vaccine and virus 
%\cite{gupta,calculator}.
\cite{gupta}.
The structure of the H3N2 hemagglutinin is shown in Figure \ref{fig0},
and the five epitopes are highlighted in color.
The quantity $p_{\rm epitope}$ 
is an accurate estimate of
influenza antigenic distance in humans.
Previous work has shown that $p_{\rm epitope}$ correlates with influenza H3N2 vaccine
effectiveness in humans with $R^2 = 0.81$ for the years 1971--2004 \cite{gupta}.
While our focus here is H3N2, other work has shown that $p_{\rm epitope}$
also correlates with influenza H1N1 vaccine effectiveness in humans 
\cite{h1n1,Huang2012}.
The $p_{\rm epitope}$ measure has been extended to 
the highly pathogenic avian influenza H5N1 viruses \cite{Peng2014}.
The $p_{\rm epitope}$ measure has additionally been extended to veterinary applications,
for example equine H3N8 vaccines \cite{Daly2013}.

In order to determine the strains to be included in the
vaccine, the emergence of new strains likely to dominate
in the human population  must be detected.
We here use the method of multidimensional scaling
to detect emerging strains.  As an example, we apply
the approach to the 2014--2015 season.
Dominant, circulating strains of influenza H3N2 in the human population
typically have been present at low frequencies for
2--3 years before fixing in the population.  While the frequencies of such
emerging strains  are low, they are high enough that
samples are collected, sequenced, and
deposited in GenBank.
Multidimensional scaling, also known as
principal component analysis \cite{Gower}, 
has been used to identify clusters of influenza from
animal model data \cite{smith}.
Thus, this method can be used to detect
an incipient dominant strain for an upcoming flu season from
sequence data alone, 
before the strain becomes dominant \cite{clustering}.
We here use this method to detect emerging strains in the 2014--2015 season.
Interestingly,
H3N2 evolves such that the reconstructed phylogenetic tree
has a distinct one-dimensional backbone \cite{Lassig2012,clustering}.

In this paper,
we show that the current  A/Texas/50/2012
vaccine is predicted not to protect against the
A/California/02/2014 strain that has emerged in the population,
consistent with recent observations \cite{WHO}.
This A/California/02/2014 strain
can be detected and predicted as a transition from the A/Texas/50/2012 strain.
The proposed summer 2015 vaccine strain is A/Switzerland/9715293/2013, which is
identical in the expressed hemagglutinin (HA1) region to
the A/California/02/2014 strain \cite{note1}.
%Old one is A/Switzerland/9715293/2013(S1S2/S2), which is on the
%company website.
%New one is A/Switzerland/9715293/2013(E4/E2). 
%These two differs in epitope A,B and D. (each site for each epitope)
Furthermore, we find that there is in 2015/2016 a transition underway from the
A/California/02/2014  cluster
to an
%A/Nebraska/04/2014 
A/New Mexico/11/2014
cluster.
The latter may be an appropriate vaccine component for next season,
because  the new A/New Mexico/11/14
cluster is emerging and appears based upon representation in the sequence database
to be displacing the A/California/02/14 cluster.

\section*{Methods}

\subsection*{The $p_{\rm epitope}$ method}

We calculate $p_{\rm epitope}$, the fraction of amino acids in the dominant epitope region of hemagglutinin that differ between the vaccine and virus \cite{gupta}. We use epitope sites as in \cite{gupta} and illustrated in Fig.\ \ref{fig1}.  For each of the five epitopes \cite{calculator,gupta}, we calculate the number of amino acid substitutions between the vaccine and virus and divide this quantity by the number of amino acids in the epitope.  The value of pepitope is defined to be the largest of these five values.

\subsection*{Identification of Vaccine Strains and Circulating Strains}

The dominant circulating influenza H3N2 strain and the vaccine strain
were determined from annual WHO reports
\cite{who04,who05,who06,who07,who2007,who08,who09,who10,Y102,Y112,who13,WHO,wer8941}.
These strains are listed in Table \ref{table0}.
In many years, the WHO report lists a preferred vaccine strain, while
the actual vaccine  is a ``like'' strain.
Additionally, in some years, different vaccines
were used in different regions.  
For each study listed in Table \ref{table0},
the vaccine strain used is listed.

\subsection*{Estimation of Vaccine Effectiveness}

Vaccine effectiveness can be quantified.  It is defined as \cite{gupta}
\begin{equation}
E = \frac{u-v}{u}
\label{effectiveness}
\end{equation}
where $u$ is the rate at which unvaccinated people
are infected with influenza, and $v$ is the rate at which
vaccinated people are infected with influenza.

The vaccine effectiveness in Eq.\ \ref{effectiveness} was calculated
from rates of infection observed in epidemiological studies.
Influenza H3N2 vaccine effectiveness values for years 1971--2004 
are from studies previously collected \cite{gupta}.
Laboratory-Confirmed
data for the years 2004--2015 were collected from the studies cited in Table \ref{table0}.
Epidemiological data from healthy adults, aged approximately 18--65, were used.
For each study,  
the total number of unvaccinated subjects, $N_u$, 
the total number of vaccinated subjects, $N_v$, 
the number of H3N2 influenza cases among the unvaccinated subjects, $n_u$,
and
the number of H3N2 influenza cases among the vaccinated subjects, $n_v$,
are known and listed in the table.
From these numbers, vaccine effectiveness was calculated
from Eq.\ \ref{effectiveness},
where $u = n_u / N_u$ and $v = n_v / N_v$.
Error bars, $\varepsilon$, on the calculated effectiveness values were obtained 
assuming binomial statistics for each data set \cite{gupta}:
$\varepsilon^2 = [\sigma_v^2/u^2 / N_v + (v/u^2)^2 \sigma_u^2 / N_u ]$,
where
$\sigma_v^2 = v (1-v)$, and
$\sigma_u^2 = u (1-u)$.

\subsection*{Virus Sequence Data in 2013 and 2014}
The evolution of the HA1 region of the H3N2 virus in
the 2013/2014 and 2014/2015 seasons was analyzed in detail.
We downloaded from GenBank the 1006 human HA1 H3N2 sequences that were
collected in 2013 and the
179 human
HA1 H3N2 sequences that were collected in 2014.

\subsection*{Sequence Data Alignment}
All sequences were aligned 
before further processing
by multialignment using Clustal Omega.
Only full length HA1 sequences of 327 amino acids were used, 
as partial sequences
were excluded in the GenBank search criterion.
Default clustering parameters in Clustal Omega were used.
There were no gaps or deletions detected
by Clustal Omega in the 2013 and 2014 sequence data.

\subsection*{Multidimensional Scaling}

Multidimensional scaling finds a reduced number of dimensions, $n$,
that best reproduce the distances between all pairs of a set of points.
In the present application, the points are HA1 sequences of length 327
amino acids, and the data were reduced to $n=2$ dimensions.
Distances between two sequences were defined as the
Hamming distance, i.e.\ the number of differing amino acids, divided
by the total length of 327.  In this way, multidimensional scaling places
the virus sequences in a reduced sequence space so that distances between
pairs of viral sequences are maintained as accurately as possible.
This low-dimensional clustering method enables one
to visualize the viruses, by finding the two best
dimensions to approximate the Hamming distances between all
clustered sequences.

%There are a greater number of sequences in recent years, than in the early 
%years of 1970--1990.  Thus, standard multidimensional scaling would
%well represent the distances between recent sequences, giving a low weight
%to the significance of the early sequences.  To remedy this uneven distribution
%of sequences per year, we here use the method of weighted multidimensional scaling.
%Each sequence is weighted by the inverse of the total number of sequences collected
%in that year. In this way, the distances between sequences in all years are given
%equal weight.

\subsection*{Gaussian Kernel Density Estimation}

The method of Gaussian kernel density estimation was used
to predict the probability density of sequences in 
the reduced sequence space identified by multidimensional scaling
\cite{clustering}.
Briefly, each sequence was represented by a Gaussian distribution
centered at the position where the sequence lies in the
reduced space.  The total estimated viral probability density was the
sum of all of these Gaussians for each virus sequence.
The weight of the Gaussian for each sequence was constant.
%equal to the weight
%of the point, i.e.\ the the inverse of the total number of sequences collected
%in that year.   
The standard deviation of the Gaussian for each sequence was specified as either
one-half, one, or three substitutions in the dominant epitope of the
virus, as discussed later.
In other words, the reconstructed probability density of the
viruses in the reduced $(x,y)$ space, as estimated by the
sequences from GenBank, was given by 
$p(x,y) \propto \sum_i \exp\{-[(x-x_i)^2 + (y-y_i)^2] / (2 \sigma^2) \} $,
where the location of virus $i$ in the reduced space is $(x_i, y_i)$,
and $\sigma$ is the standard deviation.
  In this way, a smooth estimation of the underlying
distribution of virus sequences from which the sequences deposited in
GenBank are collected is generated.

There are three criteria by which a new cluster can be judged to
determine if it will dominate in the human population in a 
future season.  First, the cluster must be evident in a
density estimation. Second,  the cluster must be growing. That is,
there must be evident selection pressure on the cluster.
Third, the cluster must be sufficiently far from the
current vaccine strain, as judged by
$p_{\rm epitope}$, for the vaccine to provide
little or no protection against the new strains.
From prior work \cite{gupta} and from the results discussed
below, 
peaks separated by more than roughly $p_{\rm epitope} = 0.19$
are sufficiently separated that protection against the virus
at one peak is expected to provide little protection against the
viruses at the other.

\section*{Results and Discussion}

\subsection*{Vaccine Effectiveness Correlates with Antigenic Distance}

\begin{figure}[tb!]
\begin{center}
\includegraphics[width=0.45\columnwidth,clip=]{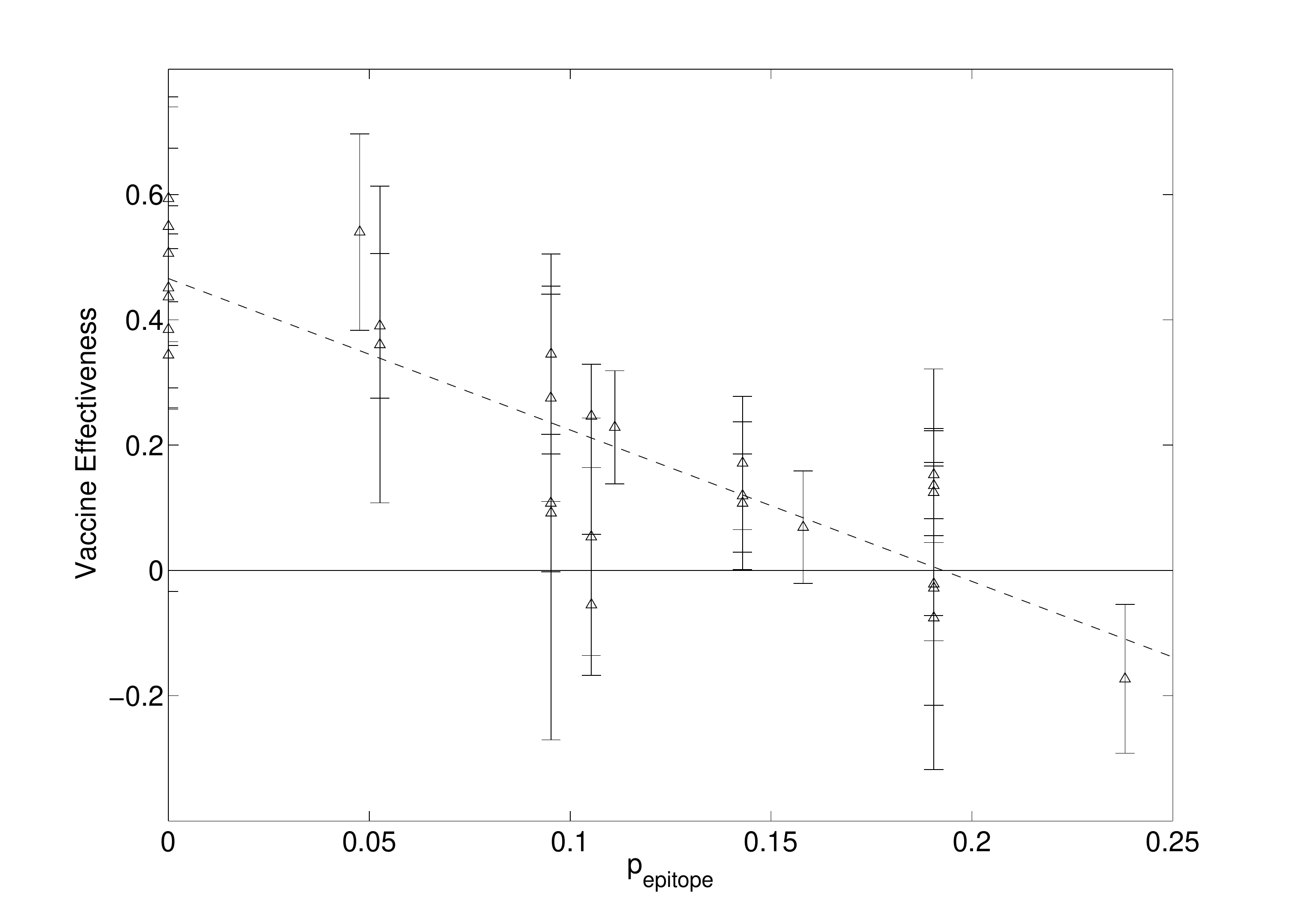}
\caption{Vaccine effectiveness in humans
as a function of the $p_{\rm epitope}$ antigenic distance.
Vaccine effectiveness values from epidemiological
studies of healthy adults, aged approximately 18--65, are
shown (triangles).
Also shown is a linear fit to the data (solid, $R^2 = 0.75$).
Vaccine effectiveness declines to zero at $p_{\rm epitope} = 0.19$ on average.
The error bars show the standard estimate of the mean of each sample point,
as discussed in the text.
\label{fig1}
}
\end{center}
\end{figure}

%We here use the  $p_{\rm epitope}$ method
%to predict vaccine effectiveness for the years 1971--2015.
%Epidemiological vaccine effectiveness
%data are taken from Ref.\ \cite{gupta} for years 1971--2004 and
%from Table \ref{table0} for years 2004--2015.
Figure \ref{fig1} shows how vaccine effectiveness decreases with
antigenic distance.
The equation for the average effectiveness (the solid line in 
Figure \ref{fig1}) is
%$E = -2.489 p_{\rm epitope}  + 0.469$.
$E = -2.417 p_{\rm epitope}  + 0.466$.

Vaccine effectiveness declines to zero
at approximately $p_{\rm epitope} > 0.19$, on average.
When the dominant epitope is A or B,  in which there are 19 or 21
amino acids respectively,
this means that vaccine effectiveness declines to zero
after roughly  4 substitutions.
%When epitope A is dominant, in which there are 19 amino acids,
%also 4 substitutions in epitope A are needed on average for
%vaccine effectiveness to decline to zero.
When the dominant epitope is C, in which there are 27 amino acids, 
the vaccine effectiveness declines to zero after roughly 5 substitutions.

Figure \ref{fig1} shows that H3N2 vaccine effectiveness in humans
correlates well with the  $p_{\rm epitope}$  measure of
antigenic distance.
In particular, the Pearson correlation coefficient
of $p_{\rm epitope}$ with H3N2 vaccine effectiveness in
humans is $R^2 = 0.75$.  
Interestingly, this correlation is nearly
the same as that previously reported for the 1971--2004 subset of years
\cite{gupta}, despite the addition of 50\% more data.
  Also of significance to note is that these
correlations with  $p_{\rm epitope}$ 
are significantly larger than those of ferret-derived distances with
vaccine effectiveness in humans, which as we will show
are $R^2 = 0.39$ or $R^2 = 0.37$ for the two most common measures.

\subsection*{Consistency of Epitopic Sites}

Analysis of HA1 sites shows that of the sites under diversifying selection
 \cite{entropy} shows, there are only 10 that by this measure should be added
 to the 130 known epitope sites \cite{gupta}.  Alternatively, of the sites
 under diversifying selection, 81\% are within the known epitope regions
 \cite{entropy}.  The 130 epitope sites that we have used nearly cover the
 surface of the head region of the HA1 protein, and this is why they are
 nearly complete.  Another recent study \cite{ref47} identified epitopes
 somewhat different from those that we use and further suggested that 
proximity to receptor binding site is a significant determinant of H3 
evolution.  This result is known to be true because the sialic acid receptor
 binding site is in epitope B, which is adjacent to epitope A, and epitopes
 A and B are the most commonly dominant epitopes over the years 
(Table \ref{table0}, 
and Table 1 of \cite{gupta}).  We note, however, that upon computing 
the
 correlation of the four epitope sites defined in \cite{ref47} with the vaccine
 effectiveness in human data considered here one finds $R^2=0.53$.  
This result is to be compared to the $R^2=0.75$ illustrated in Figure 2.

\subsection*{The Influenza A/H3N2 2014/2015 Season}
The
2014/2015 influenza vaccine contains an A/Texas/50/2012(H3N2)-like 
virus to protect against A/H3N2 viruses \cite{WHO}.
Novel viral strains detected in the human population
this year include A/Washington/18/2013, A/California/02/2014,
A/Nebraska/4/2014, and A/Switzerland/9715293/2013 \cite{wer8941}.
It should be noted that 
A/California/02/2014 and A/Switzerland/9715293/2013
are completely identical 
in the HA1 sequence that contains the HA epitopes \cite{note1}.
Table \ref{table1} shows the
$p_{\rm epitope}$ values between the vaccine strain 
and these newly-emerged strains.
The values indicate, along with Figure \ref{fig1}, that the
vaccine is unlikely to provide much protection against
these strains, since $p_{\rm epitope} > 0.19$.

\subsection*{Dynamics of Influenza Evolution}

\begin{figure}[tb!]
\begin{center}
\includegraphics [width=0.90\columnwidth,clip=] {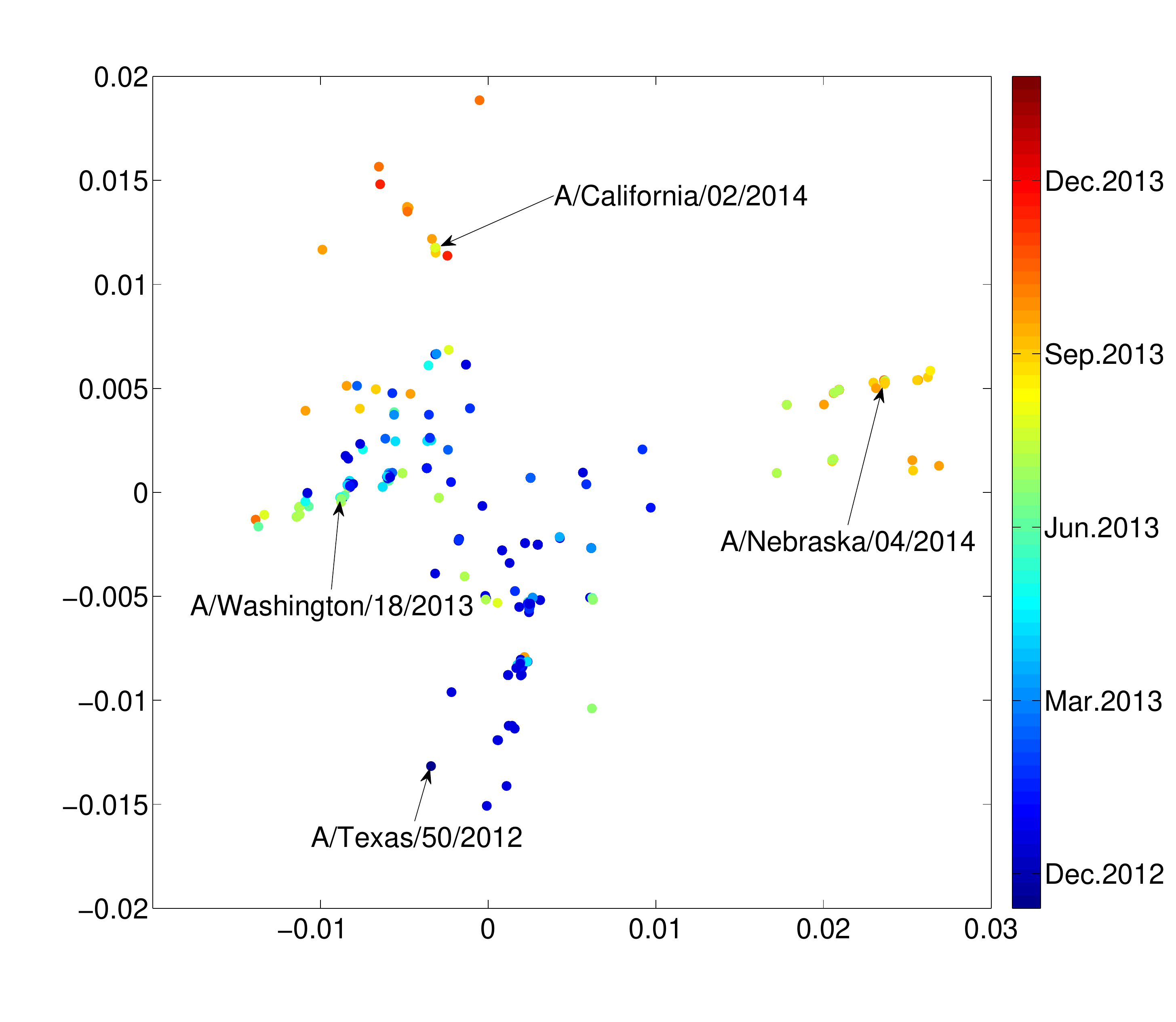}
\end{center}
\caption{Dimensional reduction of all H3N2 influenza sequences
collected from humans in 2013 and 2014 and deposited in GenBank.
Distances are normalized by the length of the HA1 sequence, 327 aa.
Dimensional reduction identifies the principal
observed substitutions, i.e.\ those
correlated with fitness of the virus, which we expect to be in 
the epitope regions.  A value of $p_{\rm epitope} = 0.19$
corresponds to a distance of $0.012$ here.
Sequences from Table \ref{table1} are labeled.
While the A/Texas/50/2012 sequence
was collected in 2012, substantially
similar strains were collected in 2013 and downloaded from GenBank.
\label{fig2}
}
\end{figure}

The strains detected in 2013 and
2014 cluster in sequence space.
While the strains are sparse in the full, high-dimensional 
sequence space, this clustering is detected
by multidimensional scaling to the two most informative dimensions,
as shown in Figure \ref{fig2}.
The novel strain A/Washington/18/2013
 emerged in 2013, followed by A/California/02/2014
and A/Nebraska/4/2014 in 2014, as shown in Figure \ref{fig2}.
The later two are sufficiently distinct from previous vaccine strains that
expected vaccine effectiveness is limited.

\begin{figure}[tb!]
\begin{center}
a) \includegraphics [width=0.45\columnwidth,clip=] {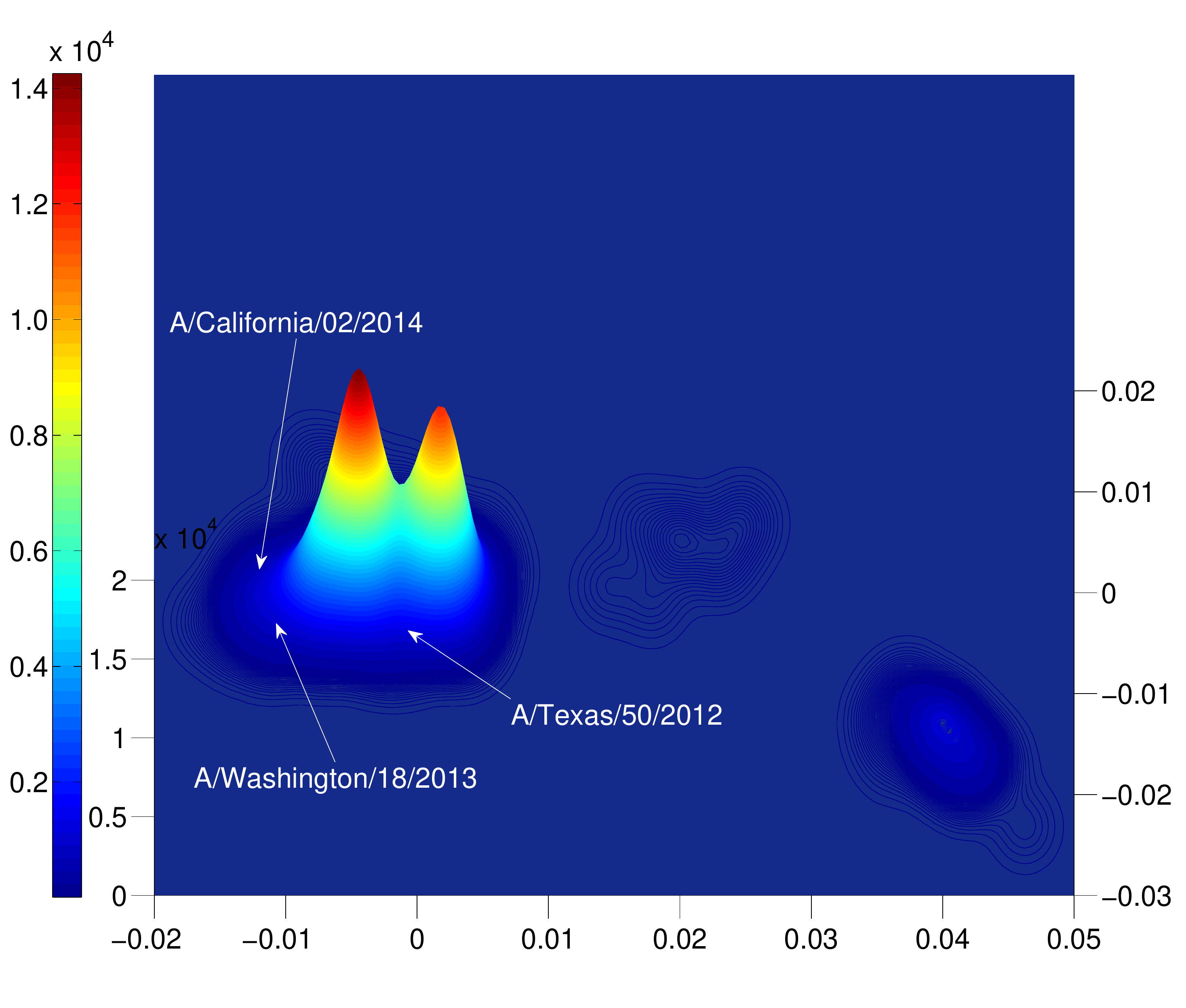}
b) \includegraphics [width=0.45\columnwidth,clip=] {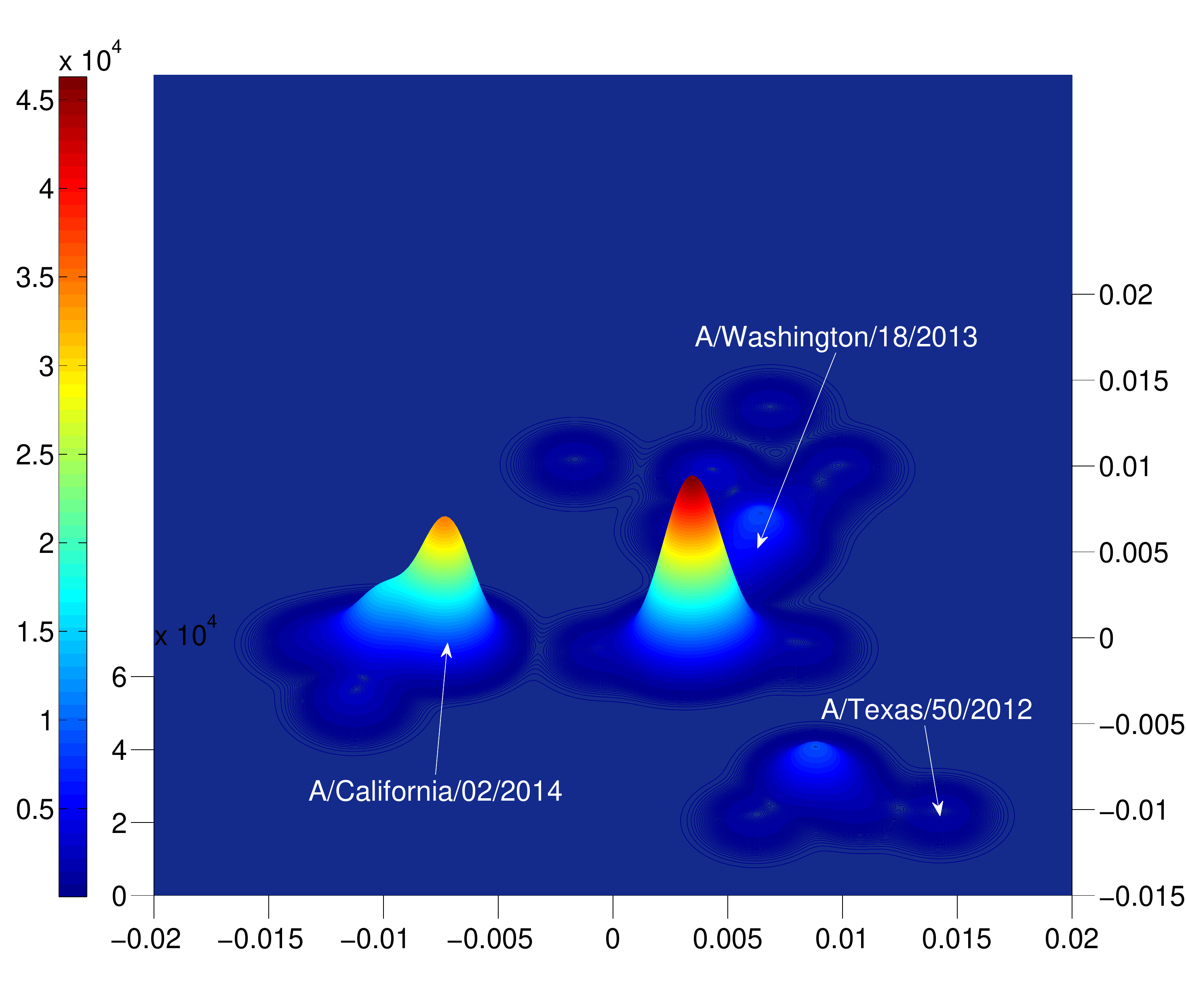}
c) \includegraphics [width=0.45\columnwidth,clip=] {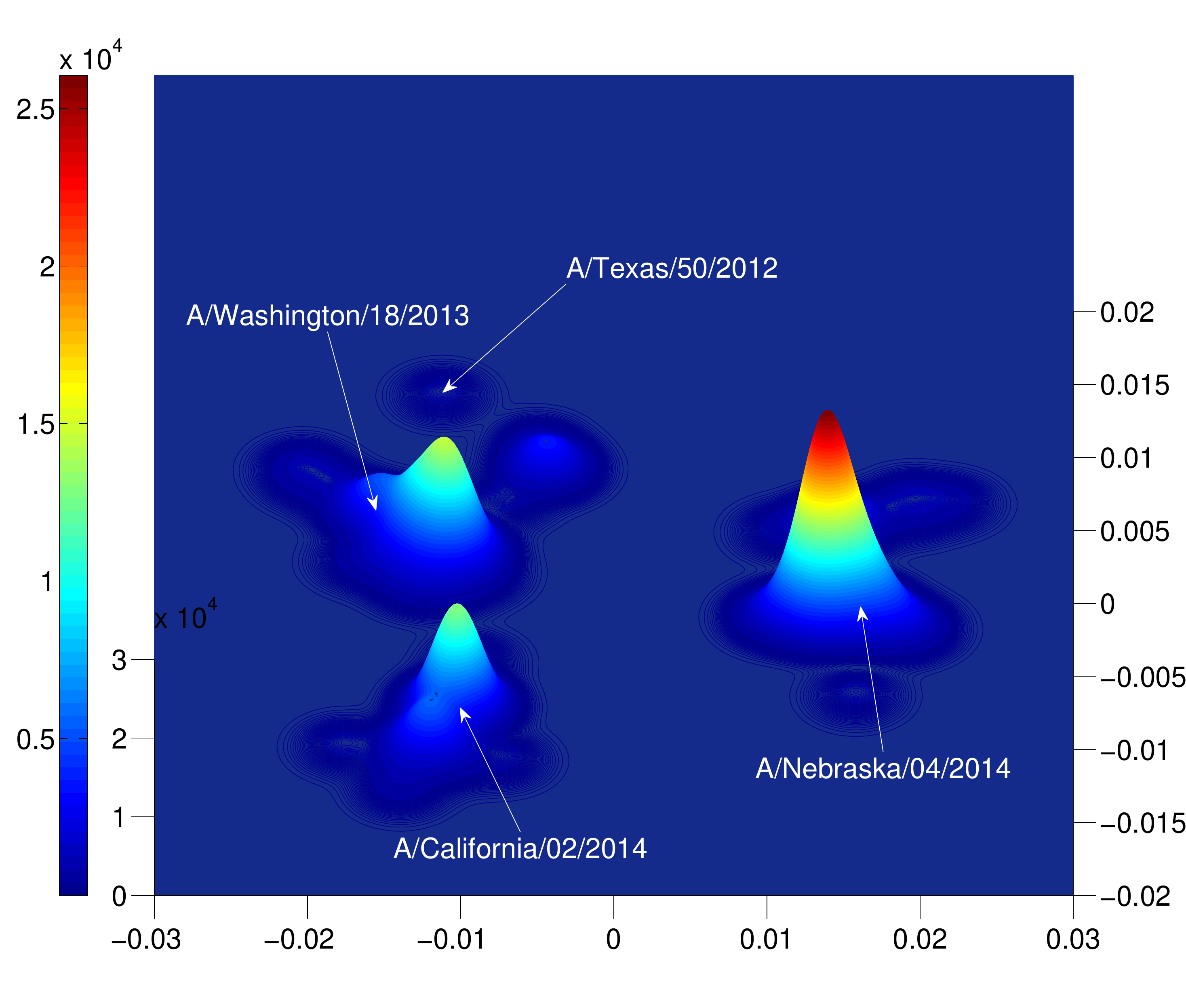}
\end{center}
\caption{Gaussian density estimation
of sequences in reduced two dimensions 
for a) all 2013 H3N2 influenza sequences in humans, 
b) those 2014 H3N2 influenza sequences in humans near
the A/Texas/50/2012 sequence,
and c) all 2014 H3N2 influenza sequences in humans.
The consensus strain of the cluster to which A/Nebraska/4/2014 
belongs is A/New Mexico/11/2014.
\label{fig3}
}
\end{figure}

Figure \ref{fig3}
is an estimate of the density distribution of the
influenza H3N2 HA1 sequences in years 2013 and 2014 
in the low-dimensional space
 provided by the multidimensional scaling.
Dimensional reduction was applied to the subset of sequences
in each subfigure \ref{fig3} a, b, or c.  Then, Gaussian kernel
density estimation was applied to estimate the distribution of sequences
in the reduced two dimensions.
Each sequence is represented by a Gaussian function with
a standard deviation of one-half substitution in the dominant epitope.

By the criteria above, A/California/02/2014(H3N2)
represented the dominant strain circulating in the human population in 2014/2015.
The time evolution in
Figure \ref{fig2}, or a comparison of Figure \ref{fig3}a with
Figure \ref{fig3}b, shows that the
A/California/02/2014 cluster emerged in 2014.
Table \ref{table1} shows that the distance of this new
cluster from the A/Texas/50/2012(egg) strain is
$p_{\rm epitope} > 0.19$, and so from Figure \ref{fig1}
the expected effectiveness of
A/Texas/50/2012(egg) against these novel A/California/02/2014-like
strains is zero.
Conversely, 
an effective vaccine for this cluster in the 2014/2015 flu season
could be 
A/California/02/2014,  or the A/Switzerland/9715293/2013 that is identical
in the HA1 region.

\subsection*{Early detection of new dominant strains}

Surprisingly, when we enlarge the region of sequence
space considered, going from Figure \ref{fig3}b to
Figure \ref{fig2} or Figure \ref{fig3}c, we find another large and growing peak 
at a distance $p_{\rm epitope} = 0.24$ from 
the A/Texas/50/2012  sequence.
This new cluster contains the
A/Nebraska/4/2014 
sequence.
The A/Nebraska/4/2014 sequence is $p_{\rm epitope} = 0.16$
from the
A/California/02/2014 sequence.
The A/Nebraska/4/2014 
 sequence appears to be dominating the
A/California/02/2014 sequence in the 2015/2016 season.
The consensus strain of this cluster to which A/Nebraska/4/2014 belongs is
A/New Mexico/11/2014.
The consensus strain minimizes the distance from all strains
in the cluster, thus maximizing expected vaccine effectiveness.
Thus, A/New Mexico/11/2014 might be a more effective
choice of vaccine for the majority of the population
in comparison to
A/Switzerland/9715293/2013 or A/California/02/2014.

\subsection*{Phylogenetic Analysis}
A systematic phylogenetic analysis of recent A/H3N2 virus HA 
nucleotide sequences has been carried out\cite{clade1,clade2}.
Briefly, phylogenetic trees were reconstructed from three reference 
sequence datasets using the maximum likelihood method \cite{clade1},
with bootstrap analyses of 500 replicates.
Dominant branches of the tree were identified with
distinct clade labels.
Analysis of the HA protein sequences showed that there were relatively 
few residue changes across all HA clades. The 2014 vaccine 
strain A/Texas/50/2012 falls into clade 3C.1, while the new 
emerging A/California/02/2014 strain falls into subclade 3C.3a. 
The A/Nebraska/4/2014 and the consensus A/New Mexico/11/2014 strains 
fall into subclade 3C.2a.
The phylogenetic analysis indicates a closer relationship 
of A/Nebraska/4/2014 or A/New Mexico/11/2014 to A/California/02/2014 than 
to A/Texas/50/2012. 
%Besides, the emerging subgroups 3C.2a and 3C.3a with low reactivity to the current vaccine are indicated. The 2014 influenza A(H3N2) viruses from South Africa (n = 34) were mainly in sublineage 3C.3 with accumulation of amino acid changes that differentiate them from the vaccine strain in 3C.1.\cite{clade2}

Note that phylogenetic methods make a number of assumptions.  For example, substitution rates at different sites are assumed to be the same and constant in time.  Due to selection, however, substitution rates are dramatically higher, at least 100x, in dominant epitope regions than in non-dominant epitope or stalk regions.  Multi-gene phylogenetic methods are inconsistent in the presence reassortment, and single-gene phylogenetic methods are inconsistent in the presence of recombination, with the former being perhaps more significant than the latter in the case of influenza.  Multidimensional scaling, on the other hand, does not make either of these assumptions.  MDS also naturally filters out neutral substitutions that are random as the dominant dimensions are identified.  Thus, MDS provides a complementary approach to the traditional phylogenetic analysis.

\subsection*{Ferret HI Analysis}
Since an analysis showing the correlations between the two standard methods
 of analyzing ferret hemagglutinin inhibition antisera assays with vaccine
 effectiveness in humans in the years 1968--2004 were $R^2=0.47$
 and $R^2=0.57$ \cite{gupta}, a number of studies have appear supporting these
 low correlations.  For example, Table 3 of \cite{ref45} shows that correlation of
 various immunogenicity parameters is higher with genetic distance than with
 HI measures of antigenic distance.  The study by Xie et.\ al further
 illustrated the limitations of relying on ferret HI data alone \cite{ref46}.
We have updated our calculation of the orrelations between the two standard methods
 of analyzing ferret hemagglutinin inhibition antisera assays with vaccine 
 effectiveness in humans to the years 1968--2015, see \cite{gupta} and the last
two columns of Table \ref{table0}.  The correlations with $d_1$ and $d_2$
are now $R^2 = 0.39$ and $R^2 = 0.37$, respectively, showing that ferret
HI studies have become even less correlated with human
vaccine effectiveness in recent years.

\section*{Conclusion}

%\section{Clustering of all sequences near Texas}

%Thus, in a principal component analysis,
%we should expect that the metric computed from the dominant components provides
%an approximation to $p_{\rm epitope}$.  The two dominant
%components in the analysis of Figure \ref{fig2} are  indeed
%sensitive to substitutions in the epitope regions of the virus.
%[TODO: show plot]
%Thus, the dimensional reduction has discovered an
%approximation to the $p_{\rm epitope}$ metric.
%Maybe discuss steps = mutations.
%Discuss why this is $p_{\rm epitope}$.
%This distance is quite close to the definition of $p_B$

In conclusion, we have shown how vaccine effectiveness can be predicted
using $p_{\rm epitope}$ values.
This method requires only sequence data, unlike 
traditional methods that require animal model data, such as
ferret HI assay experiments or post-hoc observations in humans.
Interestingly, the correlation of
$p_{\rm epitope}$ with H3N2 vaccine effectiveness in
humans is $R^2 = 0.75$,
nearly
the same as that previously reported for the 1971--2004 subset of years
\cite{gupta}, despite the addition of 50\% more data.
Significantly, the correlation of H3N2 vaccine effectiveness in humans
with  $p_{\rm epitope}$ 
is significantly larger than with ferret-derived distances,
which are $R^2 = 0.43$ or $R^2 = 0.57$ for
the two most common measures \cite{gupta}.
As an application, we estimated the 
effectiveness of the H3N2 vaccine strain of A/Texas/50/2012
against the observed A/California/02/2014 strains.

Clustering of the 2013 and 2014 sequence data confirms the
significance of the $p_{\rm epitope}$ measure. 
We showed from data through 2014 that there is
a transition underway from the A/California/02/2014  cluster
to a 
A/New Mexico/11/2014 
%A/Nebraska/04/2014 
cluster.
The consensus sequence of A/New Mexico/11/2014 
from this cluster could have been 
considered in late Winter 2015 
for inclusion among the H3N2 candidate
vaccine strains for the 2015/2016 flu season.

%\section*{Acknowledgments}
%This research was supported
%by the US National Institutes of Health under grant
%1 R01 GM 100468--01.

\clearpage

\begin{sidewaystable}
\begin{minipage}{\linewidth}
\caption{Historical vaccine strains, circulating strains, and vaccine effectivenesses.
\label{table0}}
{\centering
\begin{tabular}{l|c|c|c|c|c|c|c|c|c|c|c}
{\tiny Year}&
{\tiny Vaccine}&
{\tiny Circulating Strain}&
{\tiny Dominant Strain}&
{\tiny $p_{\rm epitope}$}&
{\tiny Vaccine}&
{\tiny $n_u$}&
{\tiny $N_u$}&
{\tiny $n_v$}&
{\tiny $N_v$}&
{\tiny $d_1$} &
{\tiny $d_2$}\\
&
{\tiny }&
{\tiny }&
{\tiny Epitope}&
{\tiny }&
{\tiny Effectiveness}&
&
&
&
&
&
\\
\hline
\hline
{\tiny 2004-2005}&
{\tiny A/Wyoming/3/2003 (AY531033)}&
{\tiny A/Fujian/411/2002 (AFG72823)}&
{\tiny B}&
{\tiny 0.095}&
{\tiny 9\% \cite{Y04}}&
{\tiny 6}&
{\tiny 40}&
{\tiny 50}&
{\tiny 367}&
{\tiny 2 \cite{d2004}} &
{\tiny 1 \cite{d2004}}
\\
{\tiny 2005-2006}&
{\tiny A/New York/55/2004 (AFM71868)}&
{\tiny A/Wisconsin/67/2005 (AFH00648)}&
{\tiny A}&
{\tiny 0.053}&
{\tiny 36\% \cite{Y05}}&
{\tiny 43}&
{\tiny 165}&
{\tiny 6}&
{\tiny 36}&
{\tiny 1 \cite{wer8109}} &
{\tiny 2 \cite{wer8109}}
\\
{\tiny 2006-2007}&
{\tiny A/Wisconsin/67/2005 (ACF54576)}&
{\tiny A/Hiroshima/52/2005 (ABX79354)}&
{\tiny A}&
{\tiny 0.105}&
{\tiny 5\% \cite{Y06}}&
{\tiny 130}&
{\tiny 406}&
{\tiny 20}&
{\tiny 66}&
{\tiny 1 \cite{d2006}}&
{\tiny 2 \cite{d2006}}
\\
{\tiny 2007}&
{\tiny A/Wisconsin/67/2005 (ACF54576)}&
{\tiny A/Wisconsin/67/2005 (AFH00648)}&
{\tiny B}&
{\tiny 0.048}&
{\tiny 54\% \cite{Y07}}&
{\tiny 74}&
{\tiny 234}&
{\tiny 8}&
{\tiny 55}&
\\
{\tiny 2008-2009}&
{\tiny A/Brisbane/10/2007 (ACI26318)}&
{\tiny A/Brisbane/10/2007 (AIU46080)}&
{\tiny }&
{\tiny 0}&
{\tiny 51\% \cite{Y08}}&
{\tiny 36}&
{\tiny 240}&
{\tiny 4}&
{\tiny 54}&
\\
{\tiny 2010-2011}&
{\tiny A/Perth/16/2009 (AHX37629)}&
{\tiny A/Victoria/208/2009 (AIU46085)}&
{\tiny A}&
{\tiny 0.053}&
{\tiny 39\% \cite{Y10,Y102}}&
{\tiny 100}&
{\tiny 991}&
{\tiny 35}&
{\tiny 569}&
{\tiny 0 \cite{d2010}}&
{\tiny 1.4 \cite{d2010}}
\\
{\tiny 2011-2012}&
{\tiny A/Perth/16/2009 (AHX37629)}&
{\tiny A/Victoria/361/2011 (AIU46088)}&
{\tiny C}&
{\tiny 0.111}&
{\tiny 23\% \cite{Y11,Y112}}&
{\tiny 335}&
{\tiny 616}&
{\tiny 47}&
{\tiny 112}&
{\tiny 1 \cite{d2011}}&
{\tiny 2.8 \cite{d2011}}
\\
{\tiny 2012-2013}&
{\tiny A/Victoria/361/2011 (AGB08328)}&
{\tiny A/Victoria/361/2011 (AIU46088)}&
{\tiny B}&
{\tiny 0.095}&
{\tiny 35\% \cite{Y12}}&
{\tiny 288}&
{\tiny 1257}&
{\tiny 15}&
{\tiny 100}&
{\tiny 5 \cite{wer8810} }&
{\tiny 4 \cite{wer8810}}
\\
{\tiny 2013-2014}&
{\tiny A/Victoria/361/2011 (AGL07159)}&
{\tiny A/Texas/50/2012 (AIE52525)}&
{\tiny B}&
{\tiny 0.190}&
{\tiny 12\% \cite{Y13}}&
{\tiny 145}&
{\tiny 476}&
{\tiny 16}&
{\tiny 60}&
{\tiny 5 \cite{wer8810} }&
{\tiny 4 \cite{wer8810}}
\\
{\tiny 2014-2015}&
{\tiny A/Texas/50/2012 (AIE52525)}&
{\tiny A/California/02/2014 (AIE09741)}&
{\tiny B}&
{\tiny 0.191}&
{\tiny 14\% \cite{Y14}}&
{\tiny 135}&
{\tiny 342}&
{\tiny 100}&
{\tiny 293}&
{\tiny 4 \cite{wer8941}} &
{\tiny 5.6 \cite{wer8941}}
\\
\end{tabular}
}
\par
\bigskip
{H3N2 influenza vaccine effectiveness in humans and
corresponding $p_{\rm epitope}$ antigenic distances
for the 2004 to 2015 seasons.
The vaccine and circulating strains 
are shown for each of the years since 2004 that
the H3N2 virus has been the predominant influenza virus and for
which vaccine effectiveness data are available. Vaccine
effectiveness values are taken from the
literature. 
%The $p_{\rm epitope}$ and $p_{\rm sequence}$
%values are calculated using eq.\  \ref{eq_epitope}
%and eq.\ \ref{eq_sequence}, respectively. 
Here $N_u$ is the total number of unvaccinated subjects, $N_v$ is the total number of
vaccinated subjects, 
$n_u$ is the number of H3N2 influenza cases 
among the unvaccinated subjects, and
$n_v$ is the number of H3N2 influenza cases 
among the vaccinated subjects.
Also shown are the distances derived from ferret HI data by the two common
measures \cite{gupta}.
}
\end{minipage}
\end{sidewaystable}

\clearpage
\begin{table}
\begin{minipage}{\linewidth}
\caption{The $p_{\rm epitope}$ distances between
the vaccine strain A/Texas/50/2012(egg) and selected 
novel strains.
 \label{table1}
}

\vspace{3mm}
\renewcommand{\baselinestretch}{1.5} \normalsize
{\scriptsize
\begin{tabular}{lcccccccc}\hline
& & \multicolumn{5}{c}{$p_i$ for each epitope $i$} & \cr\cline{3-7}  
Strain name & Collection date & A & B & C & D & E & $p_{\rm epitope}$ &
predicted effectiveness  \\

\hline
A/Texas/50/2012(cell) & 2012-04-15 & 0 & 0.0476 & 0 & 0.0244 & 0 & 0.0476 & 35\% \\
A/Washington/18/2013 & 2013-11-29 & 0.1053 & 0.1905 & 0 & 0.0244 & 0 & 0.1905 & 0\% \\
A/California/02/2014 & 2014-01-16 & 0.1579 & 0.1905 & 0 & 0.0244 & 0 & 0.1905 & 0\% \\
A/Nebraska/04/2014 & 2014-03-11 & 0.1053 & 0.2381 & 0.0370 & 0.0244 & 0.0455 & 0.2381 & 0\% \\
\hline
\end{tabular}
}
\par
\bigskip
The $p_{\rm epitope}$ distances between
the vaccine strain A/Texas/50/2012(egg) and reported
novel strains \cite{wer8941}
in 2013 and 2014.
The $p_i$ values for each epitope ($i = $ A--E), the number
of substitutions in epitope $i$ divided by the number of
amino acids in epitope $i$, are also shown.
The value of $p_{\rm epitope}$  is the largest of the $p_i$ values, and the
corresponding epitope $i$ is dominant.  Zero values indicate
no substitutions in that epitope.
\end{minipage}
\end{table}

\clearpage

\clearpage

\clearpage

\clearpage

\clearpage

\bibliography{Flu}

\end{document}